\font\tenit=cmti10
 1
 1
\font\elevenrm=cmr10 scaled\magstep 1

\documentstyle[pre,aps,epsf]{revtex}

\newcommand{\be}{\begin{equation}}
\newcommand{\ee}{\end{equation}}

\title{Nonequilibrium Kinetic Ising Models: Phase Transitions \\
and Universality Classes in One Dimension}
\author{{\sl N\'ora Menyh\'{a}rd\\ {\tenit  Research Institute for Solid 
State
Physics and Optics, \\ H-1525 Budapest,P.O.Box 49, Hungary}}\\
and\\
{\sl G\'eza \'Odor\\ {\tenit Research Institute for Technical Physics 
and
Materials Science, \\ H-1525 Budapest, P.O.Box 49, Hungary}}}

\begin{document}
\maketitle
\vskip .3cm
\vskip 0.5truecm
\date{  }
\maketitle

\begin{abstract}

{Nonequilibrium kinetic Ising models evolving under the competing effect
of spin flips at zero temperature and Kawasaki-type spin-exchange 
kinetics
at infinite
temperature $T$ are investigated here in one dimension
from the point of view of  phase transition and critical
behaviour. Branching annihilating random walks with an even number of 
offspring
(on the part of the ferromagnetic domain boundaries), is a decisive 
process
in forming the steady state of the system for a  range
 of parameters, in the family of models considered.
 A wide variety of quantities characterize the critical behaviour
 of the system.Results of computer
 simulations and of a generalized mean field theory are presented and
 discussed.}
\end{abstract}
\newpage

\section{ Introduction }
%\medskip
%\elevenrm
%\baselineskip=14pt
The Ising model is a well known static, equilibrium model. Its
dynamical generalizations, the
kinetic Ising models, were originally intended to study relaxational
processes near equilibrium states \cite{gla63,kaw72}. Glauber introduced 
the
single spin-flip kinetic Ising model, while Kawasaki  constructed
a spin-exchange version for studying the case of conserved 
magnetization.
To check ideas of dynamic critical phenomena\cite{Fer67},
 kinetic Ising models
were simple enough for producing analytical and numerical
results, especially in one dimension, where also exact solutions could 
be
obtained, and thus have proven to be a useful  ground for testing
ideas and theories \cite{rac94}.
On the other hand, when attention turned to non-equilibrium processes,
steady states, and phase transitions, kinetic Ising models were again
near at hand. Nonequilibrium kinetic Ising models, in which the steady
state is produced by kinetic processes in connection with heat
baths at different temperatures, have been widely investigated
\cite{dem85,gon87,wan88,Droz89}, and results have shown that
various phase transitions are possible under nonequilibrium
conditions, even in one dimension (1d)
(for a review see the article by R\'acz in Ref. \cite{rac96}).

Also in the field of ordering kinetics, universality is strongly 
influenced
by the conserved or non-conserved character of the order parameter.
The scaling behaviour and characteristic exponents of domain growth
are of central importance. Kinetic Ising models offer a useful
laboratory again to explore the different factors which influence these
properties \cite{Cornell}.

Most of these studies, however, have been concerned with the effects
the nonequilibrium nature of the dynamics might exert on phase 
transitions
{\it driven by  temperature}. In some examples
nonlocal dynamics can generate long-range  effective interactions
which lead to mean-field (MF)-type phase transitions (see, e.g.,  Ref. 
\cite{rac96}).

A different line of investigating nonequilibrium
phase transitions has been  via
branching annihilating random walk (BARW)  processes .
Here particles chosen at random carry out random walks
(with probability p) and annihilate pairwise on encounter.
The increase of particles in ensured
through production of $n$  offspring, with probability $1-p$.
Numerical studies have led to the conclusion that a general feature
of BARW is a transition
as a function of p between an active
stationary state with non-zero particle density and an absorbing,
inactive one in which all particles are extinct.
The parity conservation of particles is decisive
in determining the universality class of the phase transition.
In the odd-$n$ case the  universality class is that of directed
percolation (DP)\cite{DP}, while the critical behavior is different for
$n$ even \cite{jen94}. The first numerical investigation of this model
was done by Takayasu and Tretyakov \cite{Taka}.
This new universality class is often called parity conserving (PC)
(however, other authors call it the DI \cite{kim94} or BAWe class 
\cite{Bassler});
we will also refer to it by this name.
A coherent picture of this scenario is provided from a renormalization
point of view in \cite{Card97}.

The first example  of the PC transition was reported by Grassberger et
al.\cite{gra84,gra89}, who studied probabilistic cellular automata. 
These
1d models involve the processes $k\rightarrow3k$ and
$2k\rightarrow0$ ($k$ stands for kink), very similar to BARW
with $n=2$. The two-component interacting monomer-dimer model 
introduced
by Kim and Park \cite{kim94} represents a more complex system with a PC
type transition. Other representatives of this class are the 
three-species
monomer-monomer models of \cite{Bassler}, and a
generalized Domany-Kinzel SCA \cite{GDK}, which has two absorbing phases 
and an active one.

A class of general nonequilibrium kinetic Ising models
(NEKIM) with combined spin flip dynamics at $T=0$ and Kawasaki spin
exchange dynamics at $T=\infty$ has been proposed by one of
the authors \cite{men94} in which,  
for a range of parameters of the model, a PC-type transition
takes place. This model has turned out to be  very rich in several 
respects.
As compared to the Grassberger CA models, in NEKIM the rates of random
walk, annihilation and kink-production can be controlled independently.
It also offers the simplest 
possibility of studying the effect of a transition
which occurs on the level of kinks upon the behaviour of the underlying
spin system.
The present review is intended to give a summary of the results
obtained via computer simulations of the critical properties
of NEKIM \cite{MeOd95,MeOd96,MeOd97,MeOd98,DSOM}.
Dynamical scaling and a generalized mean field approximation scheme
have been applied in the interpretation.
Some new variants of NEKIM are also presented here, together with
new high-precision data for some of the critical exponents.

\section{Non-equilibrium kinetic Ising model (NEKIM)}
\subsection{ 1d Ising model}

Before going into the details of NEKIM, let us recall some
well known facts about the one-dimensional
Ising model. It is defined on a chain of length $L$;
the Hamiltonian in the presence of a magnetic field $H$
is given by ${\mathcal{H}}=3D-J\sum_{i}s_is_{i+1}- H\sum_{i}s_i$,
($s_i=\pm1$).  The model does not have a phase transition in the usual 
sense, but
when the temperature $T$ approaches zero a critical behaviour
is shown. Specifically,
$p_T=e^{-\frac{4J}{kT}}$ plays the role of $\frac{T-T_c}{T_c}$ in 1d,
and in the vicinity of $T=0$ critical
exponents can be defined as powers of $p_T$. Thus, e.g., the critical
exponent $\nu$  of the coherence length
is given  through $\xi\propto {p_T}^{-\nu}$.
Concerning statics,  from exact solutions,
keeping the leading-order terms for $T\to0$, 
the critical exponents of ($k_BT$ times) the susceptibility,
the coherence length and the magnetisation are known to be
$\gamma_{s}=\nu=1/2$, and $\beta_s=0$,
respectively. We can say that the transition is of first order.
Fisher's static scaling law 
$\gamma_s=d\nu- 2\beta_s$ is valid.

The Ising model possesses no intrinsic dynamics;
with the simplest version of Glauber kinetics (see later)
it is exactly solvable and
the dynamic critical exponent
$Z$ defined via the relaxation time of the magnetization
as $\tau=\tau_0\xi^{Z}$ has the value $Z=2$.
(Here $\tau_0$ is a characteristic time of order unity.)

Turning back to statics, if the magnetic field differs from zero
the  magnetization is also known exactly;
at $T=0$ the solution gives:
\begin{equation}
m(T=0,H)=sgn(H).
\label{eq:sgh}
\end{equation}
Moreover, for $\xi \gg 1$ and
$H/{kT} \ll 1$ the  exact solution reduces to
\begin{equation}
m\sim 2h\xi\,; \qquad h=H/k_{B}T .
\label{eq:m1}
\end{equation}
In scaling form one writes:
\begin{equation}
m\sim \xi^{-\frac{\beta_s}{\nu}}g(h\xi^{\frac{\Delta}{\nu}})
\label{eq:m2}
\end{equation}
where $\Delta$ is the static magnetic critical exponent.
Comparison of eqs. (\ref{eq:m1}) and (\ref{eq:m2}) results in
  $\beta_s=0$ and
$\Delta=\nu$ .
It is clear that the
transition is discontinuous at $H=0$ as well 
[i.e., upon changing the sign of $H$ in Eq. (\ref{eq:sgh})].
In the following the order
of limits will always be meant as: (1) $H\to 0$, and then (2) $T\to 0$.

\vglue 0.5cm
\subsection{ Definition of the model}
\medskip
\elevenrm
\baselineskip=14pt

A general form of the Glauber spin-flip transition rate in one-dimension
for spin $s_i$
sitting at site $i$ is \cite{gla63} ($s_i=\pm1$):
\begin{equation}
W_i = {\Gamma\over{2}}(1+\delta s_{i-1}s_{i+1})\left(1 -
{\gamma\over2}s_i(s_{i-1} + s_{i+1})\right)
\label{Gla}
\end{equation}
where $\gamma=\tanh{{2J}/{kT}}$ , with $J$ denoting the coupling 
constant
in
the ferromagnetic Ising Hamiltonian,
 $\Gamma$ and $\delta$ are further parameters which can, in general,
 also depend on temperature. (Usually the Glauber model is understood as 
the special
 case  $\delta=0$, $\Gamma=1$, while the Metropolis model \cite{Met}
 is obtained by choosing  $\Gamma=3/2$, $\delta=-1/3$). There are
 three independent rates:
\begin{eqnarray}
w_{same}={\frac{\Gamma}2}(1+\delta)(1-\gamma)\\
w_{oppo} ={\frac{\Gamma}2}(1+\delta)(1+\gamma)\\
w_{indif}={\frac{\Gamma}2}(1-\delta),
\label{rates}
\end {eqnarray}
where the subscripts $same$, etc., indicate the three possible
neighborhoods of a given spin($\uparrow\uparrow\uparrow, 
\downarrow\uparrow
\downarrow$ and $\uparrow\uparrow\downarrow$,respectively).
In the following $T=0$ will be taken,
thus $\gamma =1$,  $w_{same}=0$
and $\Gamma$, $\delta$ are parameters to be varied.

The Kawasaki spin-exchange transition rate of neighbouring spins
is:
\begin{equation}
w_{i,i+1}(s_i,s_{i+1})={p_{ex}\over2}(1-s_{i}s_{i+1})[1-{\gamma\over2}
(s_{i-1}s_i+
s_{i+1}s_{i+2})].
\label{Kaw}
\end{equation}

At $T=\infty$ ($\gamma=0$)  the above exchange is simply an 
unconditional
nearest neighbor exchange:
\begin{equation}
w_{i,i+1}={1\over2}p_{ex}[1-s_is_{i+1}]
\label{ex}
\end{equation}
where $p_{ex}$ is the probability of spin exchange.

The transition probabilities in Eqs. (\ref{Gla}) and (\ref{ex}) are 
responsible
for the basic elementary
processes of kinks.
Kinks separating two ferromagnetically ordered domains can carry out
random walks with probability
\begin{equation}
p_{rw}\propto 2w_{indif}={\Gamma}(1-\delta)
\end{equation}
while two kinks at neighbouring
positions will annihilate with probability
\begin{equation}
p_{an}\propto w_{oppo}={\Gamma}(1+\delta)
\end{equation}
($w_{same}$ is responsible for creation of kink
pairs inside of ordered domains at $T\not=0$).

In case of the spin exchanges, which also act only at
domain  boundaries,
the process of main importance here is that a kink can produce two
offspring at the next time step with probability
\begin{equation}
p_{k\rightarrow3k}\propto{p_{ex}}.
\end{equation}
The above-mentioned three processes compete, and it depends on
the values of the
parameters $\Gamma$, $\delta$ and $p_{ex}$ what the result of this
competition will be.
It is important to realize that  the process $k\rightarrow3k$
can  develop into propagating offspring production only if
$p_{rw}>p_{an}$, i.e., the new kinks are able to travel on the average
some lattice points away from their place of birth, and can thus avoid
immediate annihilation. It is seen from  the above definitions
that $\delta<0$ is  necessary for this to happen.
In the opposite case the only effect of the $k\rightarrow 3k$
process
on the usual Ising kinetics is to soften domain walls.
This heuristic argument is supported by simulations as well
as theoretical considerations, as will be discussed in the
next section.
\vglue .5cm

\subsection{Phase boundary of the PC transition in  NEKIM}
\medskip
\elevenrm
\baselineskip=14pt

\begin{center}
\begin{figure}[h]
\epsfxsize=90mm
\epsffile{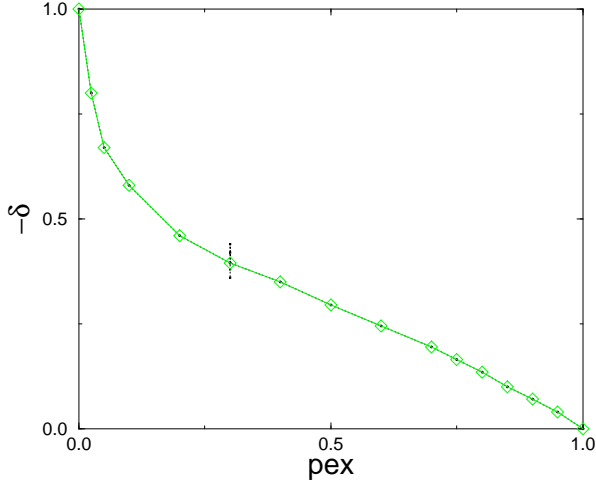}
\vspace{4mm}
\caption{Phase diagram of the two-parameter model. 
The active phase lies above the boundary.
The dotted vertical line indicates the critical point
that we investigated in greater detail.
}
\label{phasedia}
\end{figure}
\end{center}

In all of our investigations we have considered a simplified version
of the above model,
 keeping only two parameters instead of three by imposing
the normalization condition $p_{rw}+p_{an}+p_{k\rightarrow3k}=1$, 
i.e.,
\begin{equation}
 2\Gamma+p_{ex}=1.
\label{norm}
\end{equation}
In the plane of parameters $p_{ex}$ and $ 
-\delta=\frac{{p_{rw}/p_{an}}-1}
{{p_{rw}/p_{an}}+1}$ the phase diagram, obtained by
computer simulation, is as shown in Fig. \ref{phasedia}.
There is a line of second-order phase transitions,
the order parameter being the density of kinks.
The phase boundary was obtained by measuring $\rho(t)$, the density of
kinks, starting from a random initial distribution, and locating the
phase transition points via $\rho(t)\propto t^{-\alpha}$ with
$\alpha=.28\pm.01$. The lattice was typically
$L=2000$ sites; we generally averaged over 2000  independent runs.
The dotted vertical line in Fig. \ref{phasedia} indicates the critical 
point
that we investigated in greater detail;
further critical characteristics were determined along this line.

The line of phase transitions separates
two kinds of steady states reachable by the system for large times:
in the Ising phase, supposing that an even
number of kinks are present in the initial states, the system orders
 in one of the possible ferromagnetic states of all spins up or
 all spins down, while the active phase is
disordered, from the point of view of  the underlying spins, due to
the steadily growing number of kinks.

For determining the phase boundary, as well as
 other exponents discussed below, we used
 the scaling considerations of
Grassberger \cite{gra84,gra89}.
The density $\rho{(x,t,\epsilon)}$ of kinks
was assumed to follow the scaling form
\begin{equation}
\rho{(x,t,\epsilon)}\propto{t^{-\alpha}}{\phi(\epsilon{x^{1/{\nu}_\bot}},
\epsilon{t^{1/{\nu}_\Vert}})}
\label{rho}
\end{equation}
Here $\epsilon$ measures the deviation from the critical probability
at which the branching transition occurs, $\nu_{\bot}$ and
$\nu_{\Vert}$ are exponents of coherence lengths in space and time
directions, respectively.  By definition,
$\nu_{\Vert}=\nu_{\bot}Z$, where $Z$ is the dynamic critical exponent.
$\phi(a,b)$ is analytic near $a=0$ and
$b=0$. Using Eq. (\ref{rho}) the following relations
can be deduced.
When starting from a random initial state the exponent $\beta$
characterizes the growth of the average kink
density in the active phase:
\begin{equation}
\rho(\epsilon)=\lim_{t \rightarrow \infty}\rho(x,t,
{\epsilon})\propto{{\epsilon}^\beta}
\label{beta}
\end{equation}
while the decrease of density at the critical point is given by
\begin{equation}
\rho(t)\propto{t^{-\alpha}}.
\label{nt}
\end{equation}
The exponents are connected by the scaling law:
\begin{equation}
\beta=\nu_{\Vert}\alpha=\nu_{\bot}Z\alpha
\label{betascal}
\end{equation}

The phase boundary shown in Fig \ref{phasedia} was identified
using Eq. (\ref{nt}) \cite{men94}.
The initial state was random, with zero average magnetization.
At the critical point marked on the phase boundary
we recently obtained the value $\alpha=.280(5)$.
This point was chosen in
a region of the parameters $(p_{ex},\delta)$, where the effect of 
transients in
time-dependent simulations
is  small.  Critical exponents were
measured  around the point: $\Gamma=0.35$,
$\delta_{c}=-0.395\pm.001$ and
$p_{ex}=0.3$.
The deviation  from the critical
point $\epsilon$ was taken in the $\delta$-direction.
It is worth noting that in all of our simulations of the
NEKIM rule, spin-flips and spin-exchanges
alternated at each time step. Spin-flips were implemented
using two-sublattice updating, while L
exchange attempts were counted as one time-step for
exchange updating. The numerical value of the phase transition point
is sensitive to the manner of updating (thus, e.g., with sequential 
updating in both processes, $\mid{\delta_c}\mid$ decreases by more than ten
percent at the same values of $\Gamma$ and $p_{ex}$).

\begin{center}
\begin{figure}[h]
\epsfxsize=90mm
\epsffile{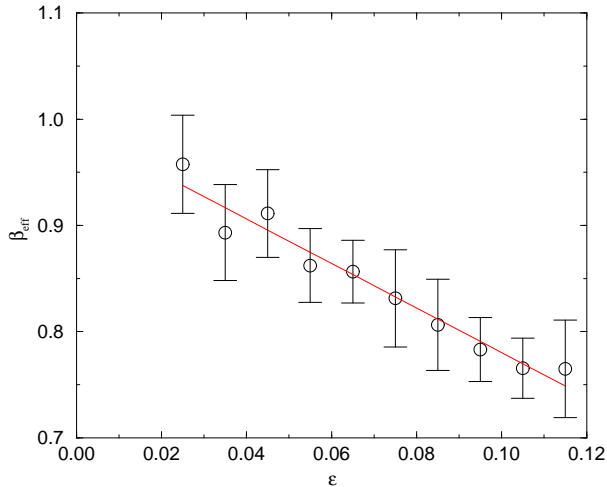}
\vspace{4mm}
\caption{Effective exponent $\beta_{eff}$ near the critical point.
Simulations were performed on a 1d NEKIM ring of size $L=24000$.
Averaging was done after the steady state was reached for
one thousand data points and $10^4$ independent samples.}
\label{betaf}
\end{figure}
\end{center}

Concerning $\beta$, we carried out quite recently a high-precision
numerical measurement; the result is shown
in Fig. \ref{betaf}. To see the corrections to scaling as well,
 we determined
the effective exponent $\beta_{eff}(\epsilon)$, which is defined as the
local slope of
$\rho(\epsilon)$ in a log-log representation between the data points 
$(i-1,i)$
\begin{equation}
\beta_{eff}(\epsilon_i) = \frac {\ln \rho(\epsilon_i) -
\ln \rho(\epsilon_{i-1})} {\ln \epsilon_i - \ln \epsilon_{i-1}} \ \ ,
\end{equation}
where $\epsilon_i = \delta_c - \delta_i$, providing an estimate
\begin{equation}
\beta = \lim_{\epsilon\to 0} \beta_{eff}(\epsilon) \,.
\end{equation}
As shown in Fig. \ref{betaf}, the effective exponent
{\em increases} slowly and tends towards
the expected PC value as $\epsilon \to 0$.
A simple linear extrapolation yields the estimate
$\beta = 0.95(2)$.

The transients mentioned above in connection with the time-dependent
simulations show up for small times and are
particularly discernible at the two ends of the phase boundary,
where the spin-flip and spin-exchange processes have very different
time scales.
For this reason, near $p_{ex}=1.0$, where we get close to
$\delta=0$ (the Glauber case), it is very difficult
to get reliable numerical estimates, even for  $\alpha$;
the exponent has been found to grow slowly with time, from a
value near zero at small times.
It is important to notice that according to Eq. (\ref{norm}), $p_{ex}=1
$
is approached together with $\Gamma\rightarrow 0$; thus
$\frac{p_{ex}}{\Gamma} \to \infty$, as otherwise $\delta_c=0$
cannot be reached.
Several  runs with different parameter values [including cases
without the restriction, Eq. (\ref{norm})], 
have been performed in this Glauber limit; all show that
the $\delta=0$ case remains
Ising-like for all values of $\Gamma$ and $p_{ex}$: 
the exponent $\alpha$ tends to $0.5$
for late times.
We therefore conjectured \cite{men94} that
the asymptote of the phase boundary
is $\delta=0$ for $p_{ex}=1.0$, and thus that
$p_{rw}>p_{an}$  is a necessary condition for a PC-type phase 
transition.

In a recent paper, using exact methods, Mussawisade, Santos and Sch\"utz 
\cite{Muss} confirmed the above conjecture. These authors study
a one-dimensional BARW model with an even number of nearest-neighbour
offspring. They start with spin kinetics; in their notation
D, $\lambda$ and $\alpha$ are the rates
of {\it spin} diffusion, annihilation and spin-exchange, respectively,
and transform to a corresponding
particle (kink) kinetics to arrive at a BARW model.
(In their notation $D=p_{RW}$, $\lambda=p_{an}$ and 
$\alpha=2p_{ex}$.)

Using standard field-theoretic techniques,
Mussawisade et al.  derived  duality relations
\begin{eqnarray}
\tilde{\lambda} = \lambda ,\\
\tilde{D}= \lambda+\alpha ,\\
\tilde{\alpha}= D-\lambda
\end{eqnarray}
which map the phase diagram onto itself in a nontrivial way
and divide the parameter space
into two distinct regions separated by the self-dual line :
\begin{equation}
D=\lambda+\alpha .
\end{equation}
The regions  mapped onto each other have, of course, the same
physical properties.
iN particular, the line $\alpha = 0$ maps onto the
line $D=\lambda$ and the fast-diffusion limit to
the limit $\alpha \to \infty$. In the fast diffusion limit the authors
find that the system undergoes
a mean-field transition, but with particle-number fluctuations
that deviate from those given by the MF approximation. So the nature of 
the phase
transition in this limit is not PC but MF.
The observation that for large $D$,
any small $\alpha$ brings the system into the active phase, together 
with
duality, predicts
a phase transition at $D=\lambda$, in the limit $\alpha \to \infty$.
The  exact result of
\cite{Muss} confirms the
conjecture  - stemming from our simulations -
on the location of the phase transition
point in the phase diagram of NEKIM in the  limit
 $p_{ex} \to 1$, which is indeed at
$\delta = 0$. The dual limit $D \to \infty$ covers the
neighbourhood of the point
$\delta = -1$, $p_{ex}=0$. This result predicts an infinite slope
of the phase boundary in this representation, at this point,
as is also apparent in Fig.\ref{phasedia}
It is important to
notice that the duality transformation does not preserve
the normalization condition of NEKIM, eq(\ref{norm}), used in all
of our numerical studies.
The self-dual line of \cite{Muss}
$D = \lambda+\alpha$,which in terms of our parametrization reads
\begin{equation}
\delta=\frac{-2p_{ex}}{1-p_{ex}}
\end{equation}
does not conserve the above-mentioned normalization condition either.
In the phase diagram of Fig. \ref{phasedia} it corresponds to a line 
which
starts at $(0,0)$ and ends at $(1/3,-1)$ in the $(p_{ex},\delta)$
plane.

Finally, in ref.\cite{Muss} a proportionality relation between the
(time-dependent) kink density
for a half-filled random initial state, and the survival
probability of two single particles in an initially otherwise empty
system has also been derived. Concerning the respective critical 
exponents
$\alpha$ and $\delta$ (the latter being defined through the critical
behaviour of the survival probability $P(t)\propto t^{-\delta}$),
they have been shown to coincide within the error of numerical
simulations in \cite{MeOd98}, as required by scaling.

\subsection{Long-range initial conditions}

The symmetry between the two extreme inital cases in time dependent
simulations (single seed versus homogeneous random state) concerning
the kink density decay inspired us
to investigate initial conditions with long-range two-point correlations
in the NEKIM model.
This procedure  was shown to cause continuously changing density
decay exponents in case of DP transitions \cite{Hayelong}.
In this way it has been possible to interpolate
continuously between the two extreme initial cases.

In the present case we have started from initial states containing an
even number of kinks to ensure kink number conservation mod 2 and
possessing kink-kink correlations of the form
$<k_ik_{i+x}> \ \propto x^{-(1-\sigma)}$ 
with $\sigma$ lying in the interval $(0,1)$. Here $k_i$ denotes a kink
at site $i$. 
These states have been generated by the same serial
algorithm as described and numerically tested in ref. \cite{Hayelong}.

\begin{figure}[h]
\epsfxsize=80mm
\centerline{\epsffile{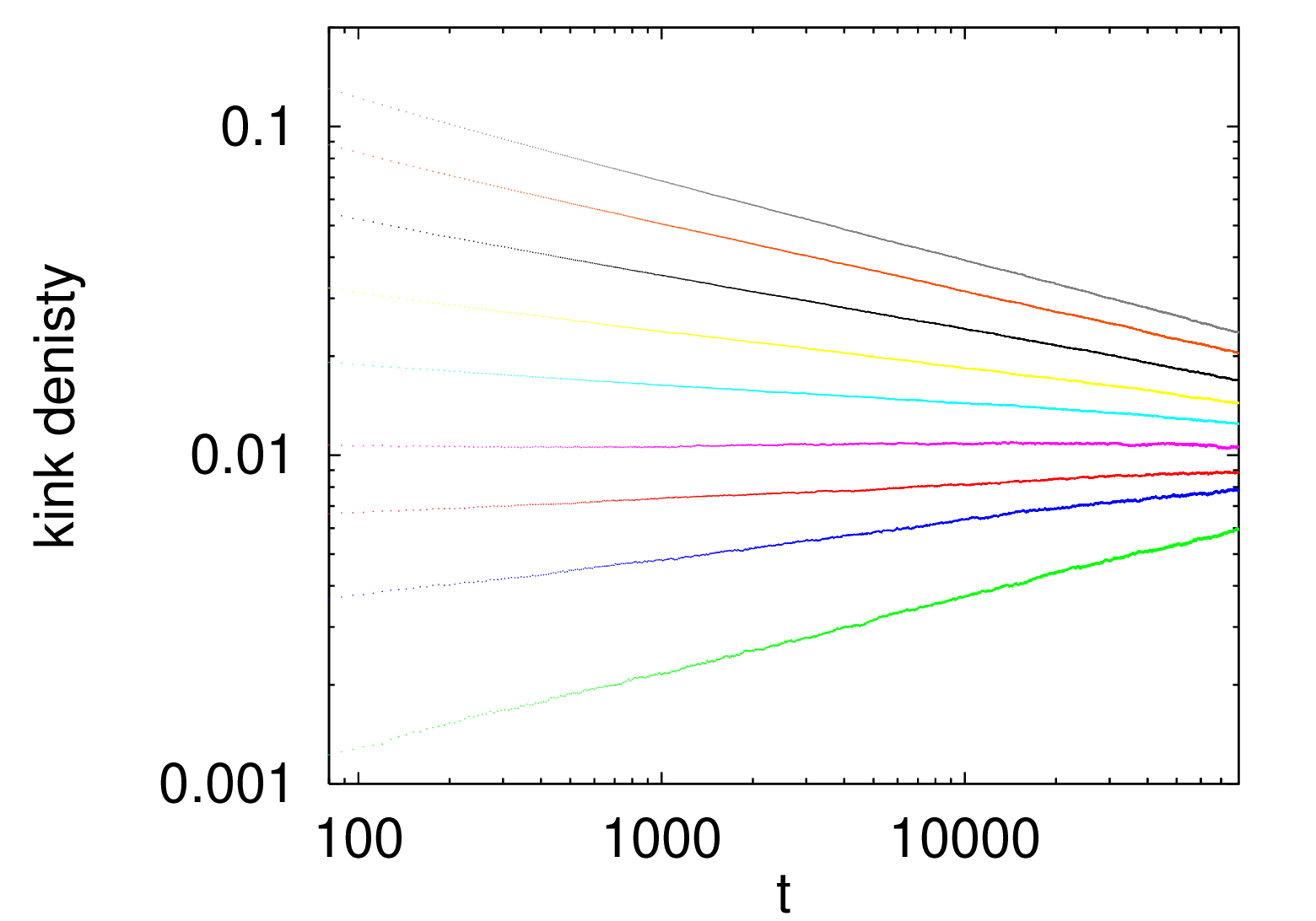}}
\vspace*{0.5cm}
\caption{$\log(\rho_{kink}(t))$ versus $\log(t)$ in NEKIM simulations 
for
initial conditions with $\sigma=0, 0.1, 0.2 ...,1$ (from bottom to 
top). }
\label{rholri}
\end{figure}

The kink density in {\it surviving} samples was measured in 
systems of $L=12000$ sites, for up to
$t_{max}=80000$ time steps; we
observe good quality of scaling over a time interval 
of three decades $(80,80000)$
(see  Fig. \ref{rholri}) \cite{Geza}.

As one can see, the kink density increases with exponent
$\alpha \sim 0.28$ for $\sigma=0$, where in principle only one pair of 
kinks
is placed on the lattice owing to the very short range correlations.
Without the restriction to surviving samples
we would have expected a constant, the above exponent corresponds
to dividing the kink density by the survival probability,$P(t)\propto
t^{-\delta}$ and the above value is in agreement with our former
simulation results for $\delta$ \cite{MeOd98}.

In the other extreme case for $\sigma=1$ the kink density decays with
with an exponent  $\alpha\sim -0.28$, again in agreement with our
expectations in case of a random initial state
\cite{men94}.
In between the exponent $\alpha$ changes linearly as a function of
$\sigma$ and changes its sign at $\sigma=0.5$. This means that the 
state
generated with $\sigma=1/2$ is very near to the situation which
can be reached for $t\to\infty$, i.e., the steady-state limit.
We mentioned at the end of the last section that the duality
transformation of ref.\cite{Muss} predicts the equality of the above two 
extremal
exponents (though it also follows from scaling).
One can make the conjecture that in
greater generality {\it the duality transformation may connect also 
cases with
initial two-point correlation exponents: $\sigma \leftrightarrow 
(1-\sigma)$.}

\subsection{Variants of the NEKIM model}

\subsubsection{ Varying exchange-range model}
Now we  generalize the original NEKIM model by
 allowing the range of the
spin-exchange to vary. Namely, Eq. (\ref{ex}) is replaced by
\begin{equation}
w_{i,i+k}={1\over2}p_{ex}[1-s_{i}s_{i+k}],
\end{equation}
 where $i$ is a randomly chosen
site and $s_i$ is allowed to exchange with $s_{i+k}$ with
probability $p_{ex}$. Site $k$ is again randomly chosen in the
interval $1\leq R$ , $R$ being thus the range of exchange.
The spin-flip part of the model will be  unchanged.
We have carried out numerical studies with this generalized
model \cite{MeOd95} in order to locate the lines of
Ising-to-active PC-type phase
transitions.
The method of updating was as before. It is worth
mentioning, that besides $k\rightarrow3k$,
the process $k\rightarrow5k$ can also occur for $R\geq3$, and
the new kink pairs are not necessarily neighbors. The character
of the phase transition line at $R>1$ is similar to that for $R=1$,
except that
the active phase extends, asymptotically, down to $\delta_c=0$.

\begin{center}
\begin{figure}[h]
\epsfxsize=80mm
\epsffile{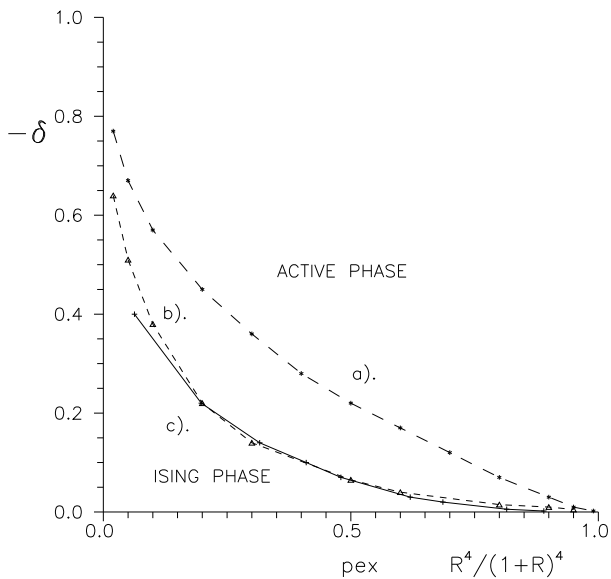}
\vspace {2mm}
\caption{Phase diagram of NEKIM for $\delta_c{(R,p_{ex})}$
and a) $R=1$, and b) $R=3$, as a function of $p_{ex}$
(with $2\Gamma=1-p_{ex}$), and for c) $\Gamma=0.35$ ($p_{ex}=0.3$)
as a function of ${(R/{1+R})}^4$ (full line). }
\label{R1}
\end{figure}
\end{center}

This is illustrated in Fig. \ref{R1}, where besides $R=1$, the
case $R=3$ is also depicted: the critical value of $-\delta_{c}$ is
shown as a function of $p_{ex}$  [curves a) and
b)]. Moreover, $-\delta_{c}$ as a function of $R$ is also
shown at constant $\Gamma=.35$ . The abscissa has been suitably
chosen to squeeze the whole (infinite) range of $R$ between $0$ and $1$
and for getting phase lines of comparable size (hence the power 4 of
$R/(1+R)$ in case of curve c)). The phase
boundaries were obtained by measuring $\rho(t)$, the density of
kinks, starting from a random initial distribution and locating the
phase transition points by $\rho(t)\propto t^{-\alpha}$ with
$\alpha=0.27\pm.04$.  We typically used a system of
$L=2000$ sites and averaged over 500  independent runs.

Besides the critical exponents we have also determined the change of
$-\delta_c$ with $1/R$
numerically at fixed $\Gamma=.35, p_{ex}=.3$ .
Over the decade of $R=4 - 40$ we have found
\begin {equation}
{-\delta_c}\approx{2.0{(1/R)}^2},
\end{equation}
which is
reminiscent of a crossover type behaviour of equilibrium and 
non-equilibrium
phase transitions \cite{rac94}, here with crossover exponent $1/2$.
It should be
noted here, that to get closer to the expected $\delta_{cMF}=0$, 
longer chains
(we used $l$ values up to $20000$) and longer runs (here up to 
$t=5\cdot10^4$)
would have been necessary.
 The former to ensure $l\gg R$ \cite{mon93}
and the latter to overcome the long transients
present in the first few decades of time steps.
In what follows we will always refer to the MF limit in
connection with $p_{ex}\rightarrow 1$ (i.e., $p_{ex}/{\Gamma}\rightarrow
\infty$), for the sake of
concreteness, but keep in mind that $R\rightarrow \infty$
can play the same role.

\subsubsection{Cellular automaton version of NEKIM}

Another generalization is the probabilistic cellular
automaton  version of NEKIM,
which consists in keeping the rates given in eqs.(\ref{rates}) and
prescribing synchronous updating. Vichniac \cite{vich} investigated
the question how well cellular automata simulate the Ising model with
the conclusion that  the unwanted feedback effects present at
finite temperatures are absent at $T=0$ and the growth of domains
is enhanced; equilibrium is reached quickly. In the language of
Glauber kinetics, this refers to the $\delta\geq 0$ case.

\begin{center}
\begin{figure}[h]
\epsfxsize=80mm
\epsffile{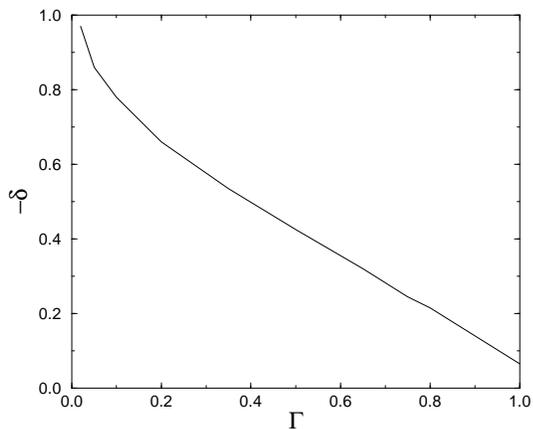}
\vspace{4mm}
\caption{ Phase diagram of the CA version of NEKIM . The phase
transition points have been determined in the same way as for
Fig.1}
\label{cadi}
\end{figure}
\end{center}

With synchronous updating, it turns out that the $T=0$ Glauber
spin-flip part of NEKIM itself leads to processes
of the type $k\to 3k$, and
for  certain values of  parameter-pairs $(\Gamma,\delta)$ with
$\delta<0$, a PC-type transition takes place.
For $\Gamma=.5$ the critical value of $\delta$ is $\delta_c=-.425(5)$
,
while e.g., for  $\Gamma=.35, \delta_c=-.535(5)$. For 
$\mid{\delta\mid}>
\mid{\delta_c}\mid$
the phase is the active one,while in the opposite case it is absorbing.
All the characteristic exponents
can be checked with a much higher speed, than for the original NEKIM.
The phase boundary of the NEKIM-CA in the $(\Gamma,-\delta)$ plane is
similar to that in Fig. \ref{phasedia} except for the highest value of 
$\Gamma=1$ $\delta_c=0$ cannot be reached; the limiting value is
$\delta_c=-.065$. For the limit  $\delta=0$ Vichniac's
considerations apply, i.e., it is an Ising phase. The limiting case
$\Gamma\to 0$ with $\delta_c\to -1$ is very hard from the point
of view of simulations,
moreover the case $\Gamma=0$ is pathological:
 the initial spin-distribution freezes in.
The corresponding phase diagram is shown in Fig. \ref{cadi}.

\subsubsection{ NEKIM in a magnetic field}
 
\begin{center}
\begin{figure}[h]
\epsfxsize=60mm
\epsffile{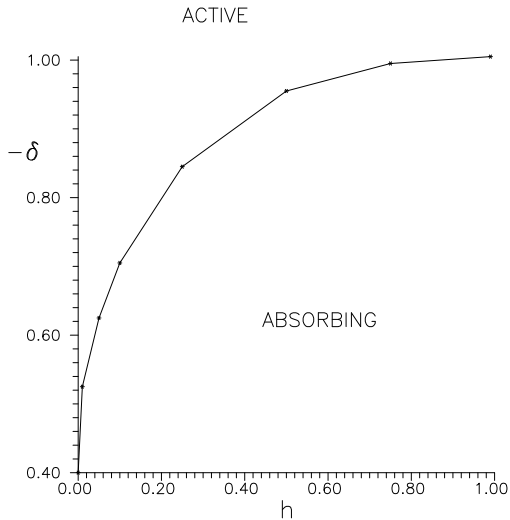}
\caption{Phase diagram of NEKIM  in the $(h,\delta)$ plane
in the presence of an external magnetic field.
Parameters of the transition probabilities: $\Gamma=.35, p_{ex}=.3$.
Naturally, the phase diagram can be drawn symmetrically for negative
values of  $h$, as well.}
\label{phasediah}
\end{figure}
\end{center}

It has been shown earlier, in the framework
of a model different from NEKIM  exhibiting a PC-type phase
transition \cite{parkh},
that upon breaking the  $Z_2$ symmetry of
the absorbing phases, DP universality is recovered.
In case of the Glauber model a similar situation arises with the
introduction of a magnetic field.
In the presence of a magnetic field
the Glauber transition rates given in Sec. 2 are modified so:

\begin{eqnarray}
w_{indif}^h& =& w_{indif}(1 - hs_i), \\
w_{oppo}^h & =& w_{oppo}(1 - hs_i), \\
         h & =&\tanh({H\over k_BT}) \ \ .
\label{hfield}
\end{eqnarray}
Fig. \ref{phasediah} shows the
phase diagram of NEKIM in the $(h,\delta)$ plane \cite{MeOd96},
starting at  the
reference PC point for $h=0$ used in this paper ($\delta_c=-.395$).

We have applied only random
initial state simulations to find the points of the
line of phase transitions (critical exponents:  $\alpha=.17(2)$,
$\beta=.26(2)$, corresponding to the DP universality class, as 
expected.)
It is seen, that with increasing field strength the critical point
is shifted to more and more negative
values of $\delta$.

Finally, it is worth mentioning a further possible varint of NEKIM
- again without
a magnetic field - in which the Kawasaki rate Eq. (\ref{Kaw})
 is considered at some finite temperature, instead of
$T=\infty$, but  keeping  $T=0$
in the Glauber dynamics. As lowering the temperature
of the spin-exchange process acts
against kink production, the result is that the active phase shrinks. 
For further details see Ref. \cite{Geza}.

\section{Generalized mean-field theory and Coherent anomaly 
extrapolation}

Using the CA version of NEKIM described in the previous section
we performed $n$-site cluster mean-field calculations (GMF) up to $n=6$,
and a coherent anomaly extrapolation (CAM), resulting in numerical estimates 
for critical exponents in fairly good agreement with those of other 
methods.
With this method we could also handle a symmetry-breaking field
that favors the spin-flips in one direction.
For this SCA model we applied the generalized mean-field technique
introduced by Gutowitz \cite{gut87} and Dickman \cite{dic88} for 
nonequilibrium
statistical systems.
This is based on the calculation of transitions of $n$-site cluster 
probabilities
$P_n(\{s_i\})$ as
\begin{equation}
P_n^{t+1}(s_1,...,s_n)={\cal F}(\{ P_m^t(s_1,...,s_m)\} 
)=P_n^t(s_1,...s_n) \,
\label{eq:defpn}
\end{equation}
where ${\cal F}$ is a function depending on the update rules and the 
last equation
refers to our solution in the steady state limit.
At the $k$-point level of approximation, correlations are
neglected for $n>k$, that is, $P_n(s_1,...,s_n)$ is expressed by using
the Bayesian extension process \cite{gut87,dlg1d}.
\begin{equation}
P_{n}(s_1,...,s_n)={\prod_{j=0}^{j=n-k}
P_k(s_{1+j},...,s_{k+j}) \over \prod_{j=1}^{j=n-k}
P_{k-1}(s_{1+j},...,s_{k-1+j})} \ .
\label{eq:bayes}
\end{equation}
The number of independent variables grows more slowly than $2^n-1$, 
owing to internal
symmetries (reflection, translation) of the update rule and the `block
probability consistency' relations:
\begin{eqnarray}
P_{n}(s_1,...,s_n) &=& \sum_{s_{n+1}} P_{n+1}
(s_1,...,s_n,s_{n+1}) \ , \nonumber \\
P_{n}(s_1,...,s_n) &=& \sum_{s_0} P_{n+1} (s_0,s_1,...,s_n)
\ . \label{eq:prob} \nonumber
\end{eqnarray}

The series of GMF approximations can now provide a basis for an
extrapolation technique. One such a method is the coherent anomaly
method (CAM) introduced by Suzuki \cite{suz86}.
According to the CAM (based on scaling) the GMF solution for a
physical variable $A$ (example the kink density) at a given level ($n$)
of approximation -- in the vicinity of the critical point, $\delta_c$ -- 
can be expressed as the product of the mean-field scaling law
multiplied by the anomaly factor $\overline{a}_n$:
\begin{equation}
A_n = \overline{a}_n \ (\delta / \delta_c^n - 1)^{\phi_{MF}} \ ,
\end{equation}
and the $n$-th approximation of critical exponents of the true singular 
behavior
($\phi_n$) can be obtained via the scaling behavior of anomaly factors:
\begin{equation}
\overline{a}_n \sim  \Delta_n^{\phi_n - \phi_{MF}} \label{anoscal}
\end{equation}
where we have used the $\Delta_n = (\delta_c / \delta_c^n -  
\delta_c^n /
\delta_c)$
invariant variable instead of $\epsilon$, that was introduced to
make the CAM results independent of using $\delta_c$ or $1 / \delta_c$ 
coupling
(\cite{kol94}). Here $\delta_c$ is the critical point estimated by
extrapolation from the sequence approximations $\delta_c^n$.
Since we can solve the GMF equations for $n \leq 6$,
we have taken into account corrections to scaling, and determined the
true exponents with the nonlinear fitting form :
\begin{equation}
\overline{a}_n = b \ \Delta_n^{\phi_n - \phi_{MF}}
+ c \ \Delta_n^{\phi_n - \phi_{MF} + 1} 
\end{equation}
where $b$ and $c$ are coefficients to be varied.

As discussed in a previous section, by duality symmetry
the $p_{ex} = 0$ axis is mapped into the $\delta=0$ axis and
the neighborhood of the
($p_{ex}=1, \delta=0$) tricritical point which  can be well 
described by
the mean-field approximations to the ($p_{ex}=0, \delta=-1$) point.
Therefore this point is also a tricritical point and the mean-field 
results
are applicable in the neighborhood.

\subsection{$Z_2$ symmetric case}

First we show the results of the GMF+CAM calculations without external 
field.
For $n=1$, we have the traditional mean-field equation for the 
magnetization:
\begin{equation}
{dm/{dt}} = -{\delta\Gamma}m(m^2-1)
\end{equation}
and for the kink density :
\begin{equation}
{d\rho/{dt}} = 2 \delta \Gamma \rho (1 - 3 \rho + 2 \rho^2 )
\end{equation}
The stable, analytic solution of the kink density exhibits a jump at 
$\delta=0$.
\[ \rho_1(\infty) = \left\{ \begin{array}{r@{\quad:\quad}l}
                  1/2 & \delta < 0 \\ 0 & \delta\ge 0
                  \end{array}\right. \]
The $n = 2$ approximation gives for the density of kinks, 
$\rho(\infty)$,
the following expression:
\begin {equation}
\rho_2(\infty) ={{{3\over 4}\,{{{ p_{rw}}}^2} + { p_{an}} - { 
p_{rw}}\,{ p_{an}}
-
     {\sqrt{{{{{1\over 16} p_{rw}}}^4} + {3\over 2}\,{{{ p_{rw}}}^2}\,{ 
p_{an}}
-
         {1\over 2}\,{{{ p_{rw}}}^3}\,{ p_{an}} + {{{ p_{an}}}^2} -
         2\,{ p_{rw}}\,{{{ p_{an}}}^2}}}}\over
   {2\,\left({1\over 2}\,{{{ p_{rw}}}^2} - { p_{rw}}\,{ p_{an}} + {{{
p_{an}}}^2}
        \right) }} \label{GMF2}
\end{equation}
for $\delta<0$. For $\delta>0$ $\rho(\infty)=0$, i.e., GMF  still
predicts a first-order transition for $\delta = 0$; the
jump size in $\rho(\infty)$ at $\delta=0$, however, decreases
monotonically with decreasing $\Gamma$, according to eq. (\ref{GMF2}).
\begin{center}
\begin{figure}[h]
\epsfxsize=150mm
\epsffile{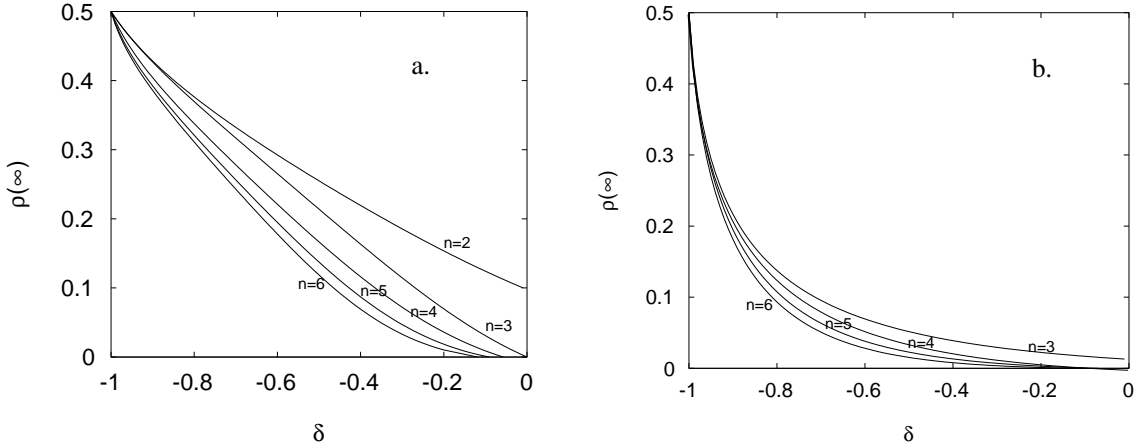}
\vspace{4mm}
\caption{$\rho(\infty)$ as a function of $\delta$ as obtained from
GMF for $n=3,..,6$ in case of a: $\Gamma=.35$, b: $\Gamma=.05$ }
\label{fig3}
\end{figure}
\end{center}
The GMF
equations can only be solved numerically for $n > 2$. We determined the 
solutions of the
$n =3,4,5,6$ approximations for the kink density at i). $\Gamma =
0.35$
(Fig.\ref{fig3} a) and of the $n=3,4,5$ approximations at ii). 
$\Gamma=0.05$
(Fig. (\ref{fig3} b)).

As we can see, the transition curves become continuous, with
negative values for $\delta_{c}^n$ ($\delta_{c}^n$ denotes
the value of $\delta$ in the $n$-th approximation for which
the corresponding $\rho(\infty)$ becomes zero).
Moreover, $\mid\delta_{c}^n\mid$ increases  with growing $n$ values. 
As increasing $n$ corresponds to decreasing mixing, i.e., decreasing 
$p_{ex}$,
the tendency shown by the above results is correct.

\begin{center}
\begin{figure}[h]
\epsfxsize=140mm
\epsffile{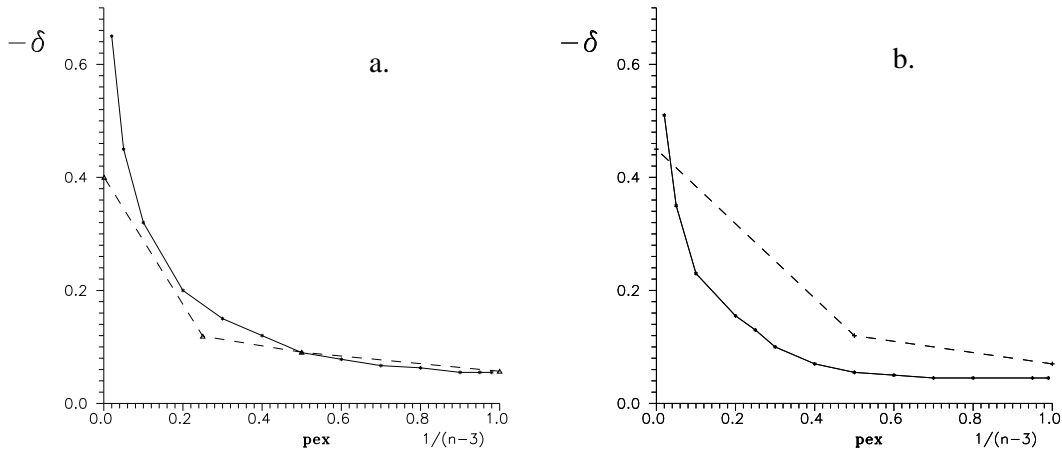}
\vspace{4mm}
\caption{ Comparison between simulation results and GMF predictions.
GMF results for $\delta_c$ are plotted as a function of $1/{(n-3)}$,
while simulation results for $\delta_c$ are depicted as a function
of $p_{ex}$ at {\it constant} $\Gamma$, with $R=3$, $\Gamma=0.35$
in case (a),
and (b) with $R=1$ for $\Gamma=0.05$.}
\label{fig5a-b}
\end{figure}
\end{center}

Figs. \ref{fig5a-b} a) and \ref{fig5a-b} b) show a quantitative -
though only tentative - comparison between the results of GMF and the
simulated NEKIM phase diagrams. The  obtained GMF data for 
$\delta_{c}^n$,
corresponding to $n=3,4,...,6$ ($\Gamma=0.35$) are depicted in
Fig. \ref{fig5a-b} a) as a function of $1/(n-3)$, together with results 
of
simulations.
The correspondence between $n$ and $p_{ex}$ was
chosen as the simplest conceivable one. (Note that $\delta_c\not=0$
is obtained first for $n=4$.) The simulated phase diagram was
obtained without requiring the normalization condition,  Eq. 
(\ref{norm}), at fixed
$\Gamma=0.35$. In this case, the $\delta_c=0$ limit, of course, is 
not reached
and a purely second-order phase transition line can be compared with the 
GMF
results
(for $n$ values where the latter also predicts a second-order 
transition).

Simulations for $R=3$ were found to lead to $\mid\delta_c\mid$
values low enough to fit GMF data. The (polynomial) extrapolation of GMF 
data to $n\rightarrow\infty$  (corresponding to $p_{ex}=0$ , i.e., 
plain spin-flip),
also shown in Fig. \ref{fig5a-b}, could be expected to approach
$\delta=-1$.
That this is not case may be due to the circumstances that upon
increasing $n$ i) GMF starts here from a first-order MF phase 
transition,
which ii) becomes second-order, and iii) GMF  should end up at the dual 
point discussed in Sec. II.C, while this symmetry is not included in the 
applied approximation procedure.
Fig. \ref{fig5a-b} (b) shows the $n=3,4,5$ results
for $\Gamma=0.05$, which are compared now with the $R=1$ simulation 
data.

The results of our GMF approximation
useful for a CAM analysis are the $n=4,5,6$ data, while the $n=3$ 
result can be taken to represent the lowest-order MF approximation
(with ${\delta_c}^{MF}=0$) for a {\it continuous} transition
(no jump in $\rho_3$). For $\delta_c$ we use a polynomial extrapolation. 

In the $n=3$ approximation the exponent $\beta=1.0064$, thus 
$\beta_{MF}\approx 1$.  Consequently, as the table below
shows, the anomaly factor
does not depend on $n$. This means, according to Eq. (\ref{anoscal}), 
that the exponent is estimated to be equal to the MF value
$\beta \simeq \beta_{MF} = 1$.
\begin{center}
\begin{table}[h]
\caption{CAM calculation results }
\begin{tabular}{|l|c|c|}
$n$ & $\Delta_n^c$   & $\overline{\rho}_n$ \\
\tableline
$4$ & $2.49043$  & $0.01083$  \\
$5$ & $1.81022$  & $0.01074$  \\
$6$ & $1.45766$  & $0.01079$  \\
\end{tabular}
\label{tablex}
\end{table}
\end{center}
\subsection{Broken $Z_2$ symmetry case}

The transition probabilities of NEKIM,
in the presence of en external magnetic field H have been given in
eqs.(27)-(29), in the previous section.
Now we extend the method for the determination of the exponent of 
the order-parameter fluctuation as well :
\begin{equation}
\chi(\epsilon) = L(<\rho^2> - <\rho>^2) \sim \epsilon^{-\gamma_n} \ \ 
.
\end{equation}
The traditional mean-field solution ($n = 1$)
results in stable solutions for the magnetization :
\begin{eqnarray}
m_1  = & - {h\over\delta}, \ \ & if \ \ \delta < 0 \ \ {\rm and} \ \ 
{h^2/
{\delta^2}}<1 \\
m_1  = & {\rm sgn}(h), \            & \  \  \  {\rm otherwise.}
\end{eqnarray}
and for the kink-concentration :
\begin{eqnarray}
\rho_1(\infty)  = & {1\over 2}(1 - ({h\over\delta})^2 ), \ \ & if\ \  
\delta<0 \ \
{\rm and} \ \ 
{h^2/{\delta^2}}<1 \\
\rho_1(\infty)  = & 0, \                                       & {\rm 
otherwise}.
\end{eqnarray}

\begin{center}
\begin{figure}[h]
\epsfxsize=150mm
\epsffile{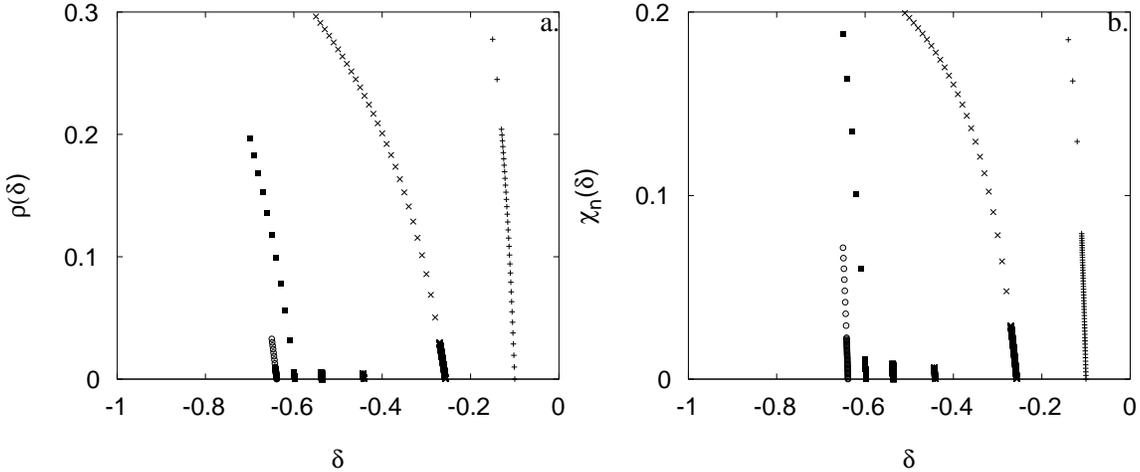}
\vspace{4mm}
\caption{The kink density (a) and (b) the second moment
of the kink density in the neighbourhood of the critical
point $\delta_{c}(h)$ for $h=0.1$.
The curves from right to left correspond to $n=1,...,6$
(level of GMF calculation). The points were determined with a resolution 
of $10^{-5}$ in $\delta$, in order to be able to extract CAM anomaly 
coefficients.
}
\label{fig12-13}
\end{figure}
\end{center}

For $n > 1$ the solutions can again only be found numerically. 
Increasing the
order of approximation, the critical point estimates $\delta_c^n$ shift 
to
more negative values similarly to the $H=0$ case.
The $\lim_{n\to\infty}\delta_c^n(h)$ values have been determined with 
quadratic
extrapolation for $h=0.01, 0.05, 0.08$, and 0.1.
The resulting curves for $\rho_n(\delta)$ and $\chi_n(\delta)$ are shown 
in Figs. \ref{fig12-13} (a) and (b), respectively, for the case of
$h=0.1$ in different orders $n$ of the GMF approximation.

Naturally, these curves exhibit a mean-field type singularity at the 
critical
point :
\begin{eqnarray}
\rho_n & \sim  \overline{\rho}_n (\delta/\delta_c^n-1)^{\beta_{MF}} \\
\chi_n & \sim  \overline{\chi}_n (\delta/\delta_c^n-1)^{-\gamma_{MF}},
\end{eqnarray}
with $\beta_{MF} = 1$ and $\gamma_{MF} = -1$.
The results of the CAM extrapolation for various $h$-values are shown in 
the table
below :

\begin{center}
\begin{table}[h]
\caption{CAM calculation results }
\begin{tabular}{|l|r|r|r|r|r|c|c|}
h       & 0.0 & 0.01 & 0.05 & 0.08 &  0.1 & DP & PC\\
\tableline
$\beta$ & 1.0 & 0.281& 0.270& 0.258& 0.285& 0.2767(4) & 0.94(1)\\
$\gamma$&     & 0.674& 0.428& 0.622& 0.551& 0.5438(13)& 0.00 \\
\end{tabular}
\end{table}
\end{center}
As to the case  $h=0$, we could not determine  the exponent
$\gamma_n$ because the low level GMF calculations resulted in
discontinuous phase transition solutions -- which we cannot use in the 
CAM extrapolation --  and so we had too few data points to achieve a
stable non-linear fitting. Higher-order GMF solutions would help,
but that requires the solution of a set of nonlinear equations
with more than $72$ independent variables.
This problem does not occur for $h \ne 0$; the above results - being 
based on the full set of approximations ($n=1,...,6$) - are fairly stable.

\section{Domain growth versus cluster growth properties; Hyperscaling}

In the field of domain-growth kinetics it has long been  accepted that
the scaling exponent of $L(t)$, the characteristic domain size,
is equal to $1/2$ if the order parameter is not conserved.
In the case of a   1d Ising spin chain of length $L$,
the structure factor at the ferromagnetic Bragg peak is defined as:
$S(0,t)=L[<m^2>-<m>^2],\,\, m=\frac{1}{L}\sum_{i}s_i$,  
($s_i=\pm1$).
If the conditions of validity of scaling are fulfilled \cite{sadiq}
then
\be
S(0,t))\propto[L(t)]^d;\qquad L(t)\propto t^x
\ee
where now $d=1$  and $x=1/2$.
Another quantity usually considered \cite{sadiq} is the excess energy
$\Delta E(t)=E(t)-E_T$ ( $E_T$ is the internal energy of a monodomain
sample at the temperature of quench), which in our case is
proportional to the kink density
$\rho(t)=\frac{1}{L}<\sum_{i}\frac{1}{2}(1-s_{i}s_{i+1})>$.
\be
\rho(t)\propto\frac{1}{L(t)}\propto t^{-y}
\label{intro}
\ee
with $y=1/2$ in the Glauber-Ising case, expressing the well known
dependence on time of
annihilating random walks ($y=\alpha$ in the notation of section 2.).
We also measured $S(t)$ at the PC point and found power-law
behaviour with an exponent $x=.575(5)$ \cite{MeOd96}.

The second characteristic growth length at the PC point is
 the cluster size, defined through the square-root of
$<R^{2}(t)>\propto t^z$.
The latter is obtained by starting either from two neighbouring kink
initial states (see e.g.\cite{jen94}), or from a single kink
\cite{gra89,men94}.
Both length-exponents
are, however connected with $Z$, the dynamic critical exponent, since
at the PC transition the only dominant length is the (time-dependent) 
correlation length. It was shown in \cite{gra89} and
\cite{gra89a} that $\frac{z}{2}=\frac{1}{Z}$
follows from scaling  for one-kink
and  two-kink initial states, respectively.
In our simulations we also found \cite{MeOd96}
that  $x=1/Z$, within numerical error, at the PC point, and thus
\be
x=z/2
\label{5.1}
\end{equation}
follows.
Exponents of cluster growth are connected by a
hyperscaling relation first established by Grassberger and de la Torre
\cite{hyper1} for the directed percolation transition.
It does not apply to the PC transition in the same form \cite{jen94},
where the dependence on the initial state (one or two kinks) manifests
itself in two cluster-growth quantities: the kink-number 
$N(t,\epsilon)\propto
t^\eta f(\epsilon t^{\frac{1}{{\nu_\bot}Z}})$ and the survival 
probability
$P(t,\epsilon)\propto t^{-\delta}g(\epsilon t^{\frac{1}{{\nu_\bot}Z}})$
(This $\delta$ has, of course, nothing to do with the parameter
$\delta$ of NEKIM).
In systems with infinitely many absorbing states, where
$\delta$ and $\eta$ depend on the density of particles in the
initial configuration, a generalised scaling relation was found
to replace the original relation,
$2\delta+\eta=z/2$, namely
\cite{mend}:
\begin{equation}
2({\beta}^{'}+{\beta}){\frac{1}{\nu_\bot {Z}}}+2\eta^{'}=z^{'}
\label{hyp}
\end{equation}
which is also applicable to the present situation.
In eq(\ref{hyp})  ${\beta}^{'}$ is defined through
$\lim_{t\to\infty}P(t,\epsilon)\propto
{\epsilon}^{{\beta}^{'}}$ and  the scaling relation 
${\beta}^{'}=\delta^{'}\nu_{\bot}{Z}$ holds.
A prime on an exponent indicates that it may depend on the initial 
state.
For the PC transition the exponent $z$ has proven (within error)
to be independent of the
initial configuration, $z^{'}=z$.

In case of a single kink initial state all samples survive thus
$\delta^{'}=0$,
and via the above scaling relation also  ${\beta}^{'}=0$. 
Eq(\ref{hyp})
 then becomes: $\frac{2\beta}{\nu_{\bot}Z}+2\eta^{'}=z$.
For $t\to\infty$, however, $\lim_{t\to\infty}N(t,\epsilon)
\propto\epsilon^{\beta}$ has to hold ( the steady state reached
can not depend on the initial state provided samples survive). Thus
using  the above scaling form for $N(t,\epsilon)$,
$\eta=
\frac{\beta}{\nu_\bot{Z}}$ follows, and the hyperscaling law can be
cast into the form:
\begin{equation}
\frac{4\beta}{\nu_{\bot}Z}=z.
\label{z}
\end{equation}

Starting with a two-kink initial state, however,
${\beta}^{'}=\beta$ and $\eta^{'}=0$  holds \cite{jen94}.
With these values Eq. (\ref{hyp}) leads  again to Eq. (\ref{z}).
As $z=2/Z$, Eq. (\ref{z}) involves only such quantities, which
are defined also when starting with
a random initial state. Using Eq. (\ref{betascal}), Eq. (\ref{z}) becomes
\begin{equation}
2y=1/Z=x
\end{equation}
In this way the factor of two between $x$ and $y$ appearing at
the PC point between  the exponents of a spin-bound and a kink-bound
quantity  (and which has been found also for a variety of
critical exponents, see \cite{MeOd96})
could be explained
as following from (hyper)scaling.

\section{Effect of the PC transition on the properties of the
underlying spin system}

Time-dependent simulations, finite-size scaling and the dynamic 
early-time MC
method \cite{li} have been applied to investigate the behaviour of the
1d {\it spin} system under the effect of the PC transition.
We found \cite{MeOd96} that within error of
the numerical studies, the dynamic critical exponent of the kinks $Z$ 
and
that of the spins $Z_c$ agree, and have the value: $Z=Z_c=1.75$. 
Thus in
comparison with the Glauber-Ising value $Z=2.0$, $Z$ decreases, as do
$\gamma$ and $\nu$:
$\gamma=\nu=.444$ (instead of 1/2).
The PC point is the endpoint
of a line of first-order phase transitions (by keeping
$p_{ex}$ and $\Gamma$ fixed and changing $\delta$ through negative 
values
to $\delta_c$), where $\beta=0$ still holds, as does
Fisher's scaling law: $\gamma=d\nu-2\beta$.
To obtain these values the PC point  has been approached from the
direction of finite temperatures of the spin-flip process,
by  varying  $p_T=e^{-4J/kT}$.

The dynamical persistency exponent $\Theta$ and 
the exponent $\lambda$\cite{MBCS96}
characterising the two-time autocorrelation function of the
total magnetization  under nonequilibrium conditions have also been
investigated numerically at the PC point.
It has been found that the PC transition has a strong effect:
the process becomes non-Markovian and the above exponents exhibit
 drastic changes as compared to the Glauber-Ising case \cite{MeOd97}.
These results together with critical exponents obtained earlier in
\cite{MeOd96} are summarized in Table III.

\begin{table}[h]
\begin{tabular}{|l|l|l|l|l|l|l|}
     & $\beta_s$& $\gamma$ & $\nu$ & Z & $\Theta\,\, $ & $\lambda\, $  
\\
\hline
Glauber-Ising& 0 & $1/2$ & $1/2$ & $2$   & $1/4$ & $1$ \\
\hline
PC &$ .00(1)$ &.444(2) &$.444(2)$ & 1.75(1) & $.67(1)$ & 1.50(2)\\
\end{tabular}
\caption{Simulation data for static and dynamic critical exponents
for NEKIM}
\end{table}
We have also investigated  the effect of critical fluctuations
at the PC point on the spreading of spins, and found
the analogue of compact directed percolation (CDP)
\cite{DoKi} to exist.
Compact directed percolation is known to appear at the endpoint
of the directed percolation critical line
of the Domany-Kinzel cellular automaton in $1+1$ dimensions\cite{DoKi}.
Equivalently, such a
transition occurs at zero temperature in a magnetic field
$h$, upon changing the sign of $h$, in the one-dimensional
Glauber-Ising model,  with
well known exponents characterising spin-cluster growth \cite{hyper2}.

\begin{center}
\begin{figure}[h]
  \centerline{\epsfysize=9cm
                   \epsfbox{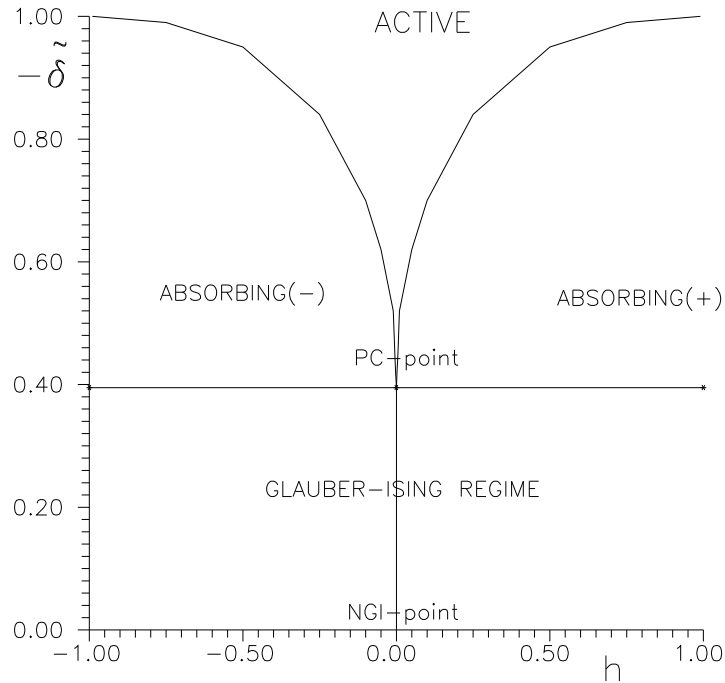}
                   \vspace*{4mm}
  }
  \caption{Phase diagram of NEKIM in the $(h,-\tilde\delta)$ plane.
The chosen PC-point is at $\tilde\delta=-0.395$.
For smaller values of  $-\tilde\delta$,
in  the Glauber-Ising regime, the vertical line connecting
the PC and the nonequilibrium Glauber Ising (NGI) points (at $h=0$)
consists of all CDP points, with its characteristic critical exponents.
The simulations around the PC transition were done for
the interval $0\leq h \leq 0.1$. The other NEKIM parameters,
in the whole plane, are: $\Gamma =0.35$ and $p_{ex}=0.3$.}
\label{hh}
\end{figure}
\end{center}

For the spreading process of a single spin $s(i)=1$ in the sea of downward spins($s(j) = -1$ for $j \ne i$), the exponents $\delta_s$, $\eta_s$ and $z_s$ are
defined at the transition point by the power-laws governing
the density of $1$'s $n_s\propto t^{\eta_s}$,
the survival probability $P_s(t)\propto
t^{-\delta_s}$, and the mean-square distance of spreading
$<R_s^2(t)>\sim t^{z_s}$.
We obtained $\eta =0$, $\delta=1/2$ and $z=1$, in accord with the
hyperscaling relation\cite{hyper2}
in the form appropriate for compact clusters:
\begin{equation}
\eta_s+\delta_s=z_s/2
\label{hl}
\end{equation}

At the PC point we found that the characteristic
exponents differ from those of the  CDP transition, as
could be expected. But basic similarities remain. 
Thus the transition which takes place upon changing
the sign of the magnetic field is of first-order and its exponents
satisfy Eq. (\ref{hl}). Accordingly it can be termed as 'compact',
and we call it the compact
parity-conserving transition
(CPC). For the phase diagram, see Fig.\ref{hh}.

On the basis of Eq. (\ref{eq:m2}),
the $t$-dependence of the magnetization in scaling form may be written 
as
\begin{equation}
m(t,h)\sim t^{-\frac{\beta_s}{\nu Z}}{\tilde g}(ht^{\frac{\Delta}{\nu 
Z}})
\label{eq:m3}
\end{equation}

We have investigated the evolution of the
nonequilibrium system from an almost perfectly magnetized initial
state (or rather an ensemble of such states) . This state is
prepared in such a way that  a single up-spin is placed in the sea
of down-spins at $L/2$.
Using the language of kinks this corresponds to the
usual initial state of two nearest neighbour kinks placed at the origin.
At the critical point the density of spins is given as
\begin{equation}
n_s(t,h)\sim t^{\eta_s}g_1(ht^{\frac{\Delta}{\nu Z}})
\label{eq:ns}
\end{equation}
for the deviation of the spin density from its initial value,
$n_s=m(t,h)-m(0)$.
The results  are summarized in Table IV.
\begin{table}[h]
\begin{tabular}{|l|l|l|l|l|l|l|}
     & $\beta_s$ & ${\beta_s}'$&$\Delta$&$\eta_s$&$\delta_s$&$z_s$ \\
\hline
NGI-CDP &$0$& $.99(2)$ & $1/2$ & $.0006(4)$ & $.500(5)$&$1(=2/Z)$ \\
\hline
CPC &$ .00(2)$&$.45(1)$&$.49(1)$ & .288(4) & $.287(3)$&$1.14(=2/Z)$ \\
\end{tabular}
\caption{Spin-cluster  critical exponents
for NEKIM in a magnetic field The NGI-CDP data are results
of simulations at the point $\delta=0$, $p_{ex}=0.3$ }
\end{table}
Concerning static exponents, only the magnetic exponent $\Delta
$ remains unchanged under the effect of the critical fluctuations.
This is not an obvious result (see, e.g., the
coherence-length exponent above).
$\beta_s{'}$, characterizing the level-off values
of the survival probability of the spin clusters
is defined through 
\begin{equation}
\lim_{t\rightarrow \infty} P_s(t,h)\propto h^{{\beta_s}'} ;
\end{equation}
this is a new static exponent connected with others via the scaling law
\begin{equation}
\beta_s{'}=\frac{\delta_s \nu Z}{\Delta}
\label{eq:scaling}
\end{equation}
which  is satisfied by the exponents obtained numerically,
within error.

\section{Damage-spreading investigations}

While damage spreading (DS) was introduced in biology \cite{Kaufman}
it has become an interesting topic in physics as well 
\cite{Creutz,Stanley,Derrida}.
The main question is if damage introduced in a dynamical system survives 
or
disappears. To investigate this the usual technique is to construct a 
pair of
initial configurations that differ at a single site,
and let them evolve with the same dynamics and external noise.
This method has been found very useful for accurate measurements of 
dynamical
exponents of equilibrium systems \cite{GrasA}.

In \cite{DSOM} we investigated the DS properties of several 1d models
exhibiting a PC-class phase transition. The order parameter that we 
measured
in simulations is the Hamming distance between the replicas:
\begin{equation}
D(t) = \left < {1\over L} \sum_{i=1}^L \vert s_i - s^,_i \vert 
\right >
\label{Dscal}
\end{equation}
where $s_i$ and the replica $s_i^,$ may denote now spin or kink 
variables.

If there is a phase transition point, the Hamming distance behaves
in a power law manner at that point. In case of initial replicas 
differing at
a single site (seed simulations) this looks like:
\begin{equation}
D(t)\propto t^{\eta} \ ,
\end{equation}
Similarly the survival probability of damage variables behaves as:
\begin{equation}
P_s(t)\propto t^{-\delta} \,
\end{equation}
and the average mean square distance of damage spreading from the center 
scales as:
\begin{equation}
R^2(t)\propto t^z \ .
\end{equation}

In case of the NEKIM model we have investigated the DS on both the spin
and kink levels. We found that the damage-spreading point
coincides with the critical point, and that the kink-DS transition 
belongs to the
PC class.
The spin damage exhibits a discontinuous phase
transition, with compact clusters and PC-like spreading exponents.
The static exponents determined by finite-size scaling are consistent
with those of spins of the NEKIM model at the PC transition point and
the generalised hyperscaling law is satisfied.

By inspecting our large data set for various models exhibiting
multiple absorbing states, we are led to a final conjecture:
{\it BAWe dynamics and $Z_2$ symmetry of the
absorbing states together form the necessary condition to have a
PC-class transition.}

\section{Summary}

The present paper has been devoted to reviewing most of  the authors'
investigations of a
nonequilibrium kinetic Ising model (NEKIM).

NEKIM has proven to be a good testing ground
for ideas, especially connected with dynamical scaling,
in the field of nonequilibrium phase transitions.
The field covered here belongs to the class of absorbing-state
phase transitions which are usually treated on the level of
particles, the best known examples being  branching annihilating
random walk models. The phase transition
on the level of kinks in the Ising system belongs
the the so-called parity-conserving universality class. The NEKIM model
offers the possibility of revealing and clarifying features and
properties which take place on the level of the underlying spin
system as well. The introduction of a magnetic field in the
Ising problem has also led to understand different features,
e.g., the change of the universality class from PC to DP
in the critical behavior at (and in the vicinity of) the
phase transition of kinks.

Investigations of various critical properties have been
carried out mainly with the help of numerical simulations
(time-dependent simulations, finite-size scaling and dynamic early-time 
MC
methods have been applied). In addition, a generalized mean-field
approximation was used, particularly in the neighborhood
of a limiting
situation of the phase diagram of NEKIM, where the usually second-order
phase transition line ends at a MF-type first-order point. Some of the
critical exponents, especially that of the order parameter, could
be predicted to high accuracy in this way.

While from the side of numerics a lot of information has accumulated 
over
the last five years concerning the PC universality class, further
investigations are needed at the microscopic level.

\bigskip
{\bf Acknowledgements}\\
The authors would like to thank the Hungarian research fund OTKA ( Nos.
T-23791, T-025286 and T-023552) for
support during this study. One of us (N.M.) would like to acknowledge
support by SFB341 of the Deutsche Forschungsgemeinschaft, during her
stay in K\"oln, where this  work was completed.
G. \'O  acknowledges support from Hungarian research fund
B\'olyai (No. BO/00142/99) as well.
The simulations were performed partially on the FUJITSU AP-1000 and AP-3000
supercomputers and Aspex's System-V parallel processing system
(www.aspex.co.uk).

\end{document}